\documentclass[aip,twocolumn,reprint,a4paper]{revtex4-1}

\usepackage{graphicx}

\begin{document}

\title{Efficient demagnetization protocol for the artificial triangular spin ice}

\author{J.H. Rodrigues}
\affiliation{Departamento de F\'{i}sica, Universidade Federal de Vi\c{c}osa, 36570-000 - Vi\c{c}osa - Minas Gerais - Brazil.}
\affiliation{Departamento de F\'{i}sica, ICEx, Universidade Federal de Minas Gerais, 31270-901 - Belo Horizonte - Minas Gerais - Brazil.}

\author{L. A. S. M\'{o}l}
\email{lucasmol@fisica.ufmg.br}
\homepage{https://sites.google.com/site/lucasmol2}
\affiliation{Departamento de F\'{i}sica, Universidade Federal de Vi\c{c}osa, 36570-000 - Vi\c{c}osa - Minas Gerais - Brazil.}
\affiliation{Departamento de F\'{i}sica, ICEx, Universidade Federal de Minas Gerais, 31270-901 - Belo Horizonte - Minas Gerais - Brazil.}

\author{W. A. Moura-Melo}
\email{winder@ufv.br}
\homepage{https://sites.google.com/site/wamouramelo/}
\affiliation{Departamento de F\'{i}sica, Universidade Federal de Vi\c{c}osa, 36570-000 - Vi\c{c}osa - Minas Gerais - Brazil.}

\author{A. R. Pereira}
\email{apereira@ufv.br}
\homepage{https://sites.google.com/site/quantumafra/home}
\affiliation{Departamento de F\'{i}sica, Universidade Federal de Vi\c{c}osa, 36570-000 - Vi\c{c}osa - Minas Gerais - Brazil.}

\date{\today}

\begin{abstract}
In this work we study demagnetization protocols for an artificial spin ice in a triangular geometry. Our results show that a simple hysteresis-like process is very efficient in driving the system to its ground state, even for a relatively strong disorder in the system, confirming previous expectations. In addition, transitions between the magnetized state and the ground state were observed to be mediated by the creation and propagation of vertices that behave like magnetic monopoles pseudo-particles. This is an important step towards a more detailed experimental study of monopole-like excitations in artificial spin ice systems.
\end{abstract}

\maketitle

During the last years, the study of frustrated magnetic systems is undergoing a great revolution. With the development of modern experimental techniques, it is nowadays relatively easy to build arrays of magnetic nanoparticles in such a way that the interactions between them can be controlled by changing the lattice geometry. This possibility opened the door to propose and realize systems with desired properties, with artificial spin ice (ASI) systems~\cite{Wang06,Qi08,Schumann10,Ladak10,Li10,Ke08,Zhang12,Mol12,Nascimento12} being the one of the greatest examples.

In ASI, a lattice of elongated magnetic nanoparticles, designed to be frustrated, is built in such a way that not all pairwise interactions between islands in a vertex can be satisfied simultaneously. In general these systems are designed to mimic the frustration present in crystalline spin ices~\cite{Harris97,Ryzhkin05,Castelnovo08}; they also support collective excitations that are expected to behave as magnetic monopoles~\cite{Nambu,Mol09,Mol10,Silva12,Mol12,Silva12b}. However, in many of the possible ASI realizations, the unbalance between dipolar interactions among collinear and non-collinear islands and the long-range character of interactions enforce a non-degenerated ground state. Since monopole excitations are expected to live in a ground state background~\cite{Mol09,Mol12}, the experimental achievement of the ground state is an important step towards a better understanding of the interesting properties of these systems. Nevertheless, only a relative success in driving these systems to their ground states was obtained so far; indeed, it is usually obtained when as-grown~\cite{Morgan11,Nisoli12} or thermal~\cite{Porro13} systems are investigated and domains in the ground state are found. The difficulty to access the ground state by other means, e.g., by using external magnetic fields, is attributed to a dynamical bottleneck~\cite{Budrikis10,Budrikis11,Budrikis12,Budrikis12b,Budrikis12c,Wysin12}.

In a recent paper~\cite{Mol12} some of us proposed that an artificial spin ice in a triangular geometry (ATSI) have also, as low energy excitations, a kind of Nambu monopole pseudo-particle as it occurs with the artificial square spin ice. An expected advantage of ATSI is the apparently facility to access its ground state, allowing a more detailed study of the properties of monopole excitations.
In order to verify this hypothesis we have carefully analyzed this possibility in this paper by using computer simulations. Our results indicate that the ground state of the artificial triangular spin ice can be easily obtained by using a simple hysteresis-like process even if the system contains a relatively strong disorder.

\begin{figure}
\includegraphics[scale=1]{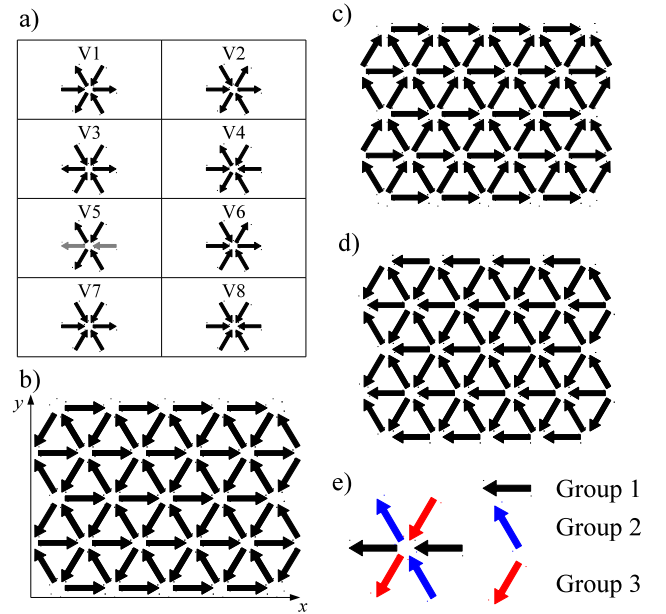}
\caption{\label{vertices} (a) The eight different vertex topologies of the triangular spin ice in order of increasing energy. In (b), (c) and (d) are shown the configurations of the ground state, of a state magnetized along the diagonal direction and of the completely magnetized state, respectively. e) The three different groups of spins.}
\end{figure}

As discussed in Ref.~\onlinecite{Mol12}, the $64$ possible vertex configurations of the ATSI (see in Fig.~\ref{vertices}(a)) can be classified in $8$ different topologies according to their energy. In a completely magnetized state, all vertices are in topology $V5$, which satisfies the $3$-in/$3$-out rule, but such a state is much more energetic than the ground state topology $V1$ (see Figs.~\ref{vertices}(b) and (c)). By considering only single spin flip processes, the simplest way to turn a $V5$ vertex into a $V1$ vertex is obtained by flipping one of the gray spins in Fig~\ref{vertices}(a-5), transforming it a $V4$ vertex. Then the other gray spin can be flipped in such a way that topology $V1$ is achieved. Should be noted, however, that by flipping any of the black spins in Fig~\ref{vertices}(a-5) a $V6$ vertex is obtained, increasing the system's energy. In this way, it seems that the completely magnetized state can be seen as a saddle point in phase space. Of course, since long-range dipolar interactions are present, this simple vertex analysis may not describe the system correctly and, therefore, a more detailed study is needed.

Here we model the ATSI by considering punctual Ising-like dipoles as in Ref.~\onlinecite{Mol12}. To investigate the hysteresis-like process, we used the same methodology employed by Budrikis et al.~\cite{Budrikis10,Budrikis11,Budrikis12,Budrikis12b,Budrikis12c} to study demagnetization protocols for the artificial square spin ice. The main idea behind this method is that a spin is flipped if $\vec{B}_i^{tot} \cdot \vec{S_i} < - h_i.$
In this expression $\vec{B}_i^{tot}=\vec{B}_i^{lat}+\vec{B}_i^{ext}$ is the total magnetic field at spin $i$ ($\vec{B}_i^{lat}$ is the dipolar magnetic field produced by all spins of the lattice at the position where spin $i$ is placed and $\vec{B}_i^{ext}$ is the external applied magnetic field at the same position); $\vec{S}_i$ is the spin at site $i$ and $h_i$ is a switching constant field (with different values for different sites $i$).

In this work, we have considered that all possible effects of imperfections during the system production~\cite{Budrikis12b,Kholi11,Pollard12,Daunheimer11} can be reproduced by considering only a disorder in the values of of the ``switching field'' $h_i$~\cite{Budrikis12c}. Then $h_i$ is obtained by a Gaussian distribution centered in $h_c$ with variance $\sigma$ in a sample and averages are accumulated for different realizations of disorder. 
In order to associate our results to experiments, we considered Permalloy nanoislands, whose saturation magnetization is about $8\times 10^7 A/m $ for volume about  $7.5\times 10^5 nm^3$ ($300\times 100 \times 25 nm$), giving a magnetic moment $\mu \sim 6\times 10^{-16} Am^2$. Thus, for a lattice spacing ($a$) of $500 nm$, our energy scale is $D=\mu_0 \mu^2/4\pi a^3 \sim 3\times 10^{-19}J$  and our magnetic field scale is given by $D/\mu \sim 0.5 mT $. Of course, these values can be re-scaled to meet experiments. From recent experiments~\cite{Phatak12}, in our simulations we have chosen for the switching field $h_c=50 mT$ and $0\leq \sigma \leq 10 mT$ ($\sim 0.2  h_c$).

\begin{figure}
\includegraphics[scale=1]{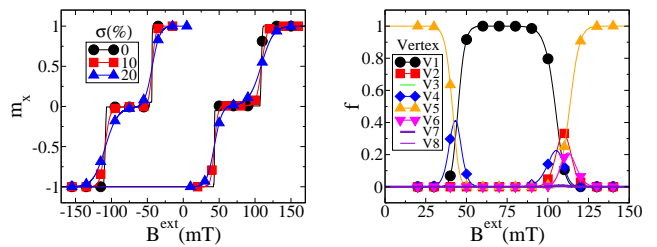}
\caption{\label{theta0} (color online) Left: $x$-component of the magnetization as a function of the external magnetic field in a hysteresis-like process for different strengths of disorder. Right: Vertices fraction for $B^{ext} > 0$ portion of the hysteresis curve for a system with $\sigma=0.1 h_c = 5 mT$. }
\end{figure}

We start by showing the results for a hysteresis-like process for a lattice with linear dimension $L=30a$. In our calculations $\vec{B}^{ext}$ makes an angle $\theta$, up to $10$ degrees, with the $x$-direction. First, considering $\theta =0$, we show in Fig~\ref{theta0} (left) the $x$-component of the magnetization ($m_x$) as a function of the external field $B^{ext}$ for different values of $\sigma$. As can be seen, there is a large plateau at $m_x=0$ for all values of $\sigma$. In Fig.~\ref{theta0} (right) we show the fraction of each vertex topology as a function of $B^{ext}$ 
for $\sigma=0.1 h_c = 5$ mT. 
In the first transition, $V5$ vertices give rise to $V4$ vertices. After that, the $V4$ vertices are destroyed to form $V1$ vertices in such a way that, for $B^{ext}\approx 60 mT $, all vertices are in topology $V1$, i.e., the system reaches its ground state. This process may be summarized in the following way: $V5 \to V4 \to V1$ as predicted in Ref.~\onlinecite{Mol12} and it is mediated by the creation and propagation of vertices that do not satisfy the $3$-in/$3$-out rule ($V4$ vertices). Mainly, $V4$ vertices are created at the borders by the inversion of spins belonging to the group $1$ shown in Fig.~\ref{vertices}(e), propagating inside the system, leaving behind them a sequence of $V1$ vertices. This process occurs until a $V4$ vertex finds another $V4$ vertex to annihilate and give rise to a $V1$ vertex. Of course, inner spins with smaller switching field are more susceptible to flip, in such a way that many pairs of $V4$ vertices are created inside the sample. We also observed that, only after all spins belonging to group $1$ inside the system change their direction, the outer spins, at the lattice's edge, flip. Another interesting observation is that there is a tendency of spin's lines be inverted in an order that favors the existence of an antiferromagnetic intermediary state. In addition, the main effect of increasing disorder is to favor flips of spins with smaller switching field first and to prevent flips of spins with higher switching field, similarly to what happens in Ref.~\onlinecite{Mengotti10}. 

In the second transition of Fig.~\ref{theta0} (right), the system completes the magnetization reversion. In this process the spins belonging to groups $2$ and $3$ shown in Fig.~\ref{vertices}(e) must flip, since all spins belonging to group $1$ have flipped in the first transition. In most part, this process is mediated by the creation and propagation of $V4$ vertices, following a sawtooth path, leaving behind them, $V2$ vertices. Then, another vertex that does not satisfies the $3$-in/$3$-out rule, $V6$, is created and also propagates following a sawtooth path of $V2$ vertices; they leave behind them $V5$ vertices. In summary, one has $V1 \to V4 \to V2$ and then $V2 \to V6 \to V5$. However, the presence of increasing disorder favors the appearance of $V7$ vertices (specially in small systems) as well as it also favors the creation of almost isolated $V4$ and $V6$ vertices in the inner region of the lattice. Moreover, $V7$ vertices are quite common during reversal, specially in the out-of-equilibrium processes that occur when the external field changes, which explains its low fraction in the equilibrium reversal curve shown here. To better illustrate these transitions, in the supplementary material\cite{supplementary}, there is an animation made from a simulation of a small lattice.

\begin{figure}
\includegraphics[scale=1]{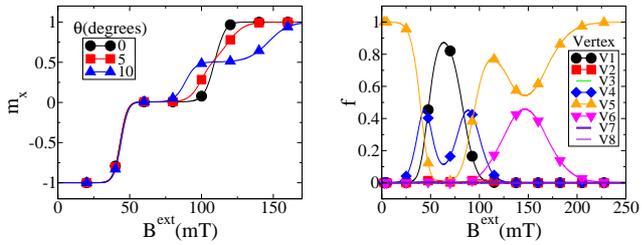}
\caption{\label{theta10} (color online) Left: $x$-component of the magnetization as a function of the external magnetic field for the $B^{ext} > 0$ portion of the hysteresis-like process and for different angles between the $x$-axis and the external field direction ($\theta$). Right: Vertices fraction for the $B^{ext} > 0$ portion of the hysteresis curve for a system with $\sigma=0.2 h_c = 10$ mT and an angle $\theta = 10^\circ$.  }
\end{figure}

Now, considering $\theta > 0$ and $\sigma=0.2h_c=10 mT$, the main difference observed is the appearance of a second plateau in the hysteresis curve, as shown in Fig.~\ref{theta10}. Its appearance can be easily explained by the fact that the external field induces the flipping of spins belonging to group $3$ (see Fig.~\ref{vertices}(e)), in such a way that the configuration of Fig.~\ref{vertices}(c) is obtained. Note however that, the ground-state plateau at $m_x=0$ is still present, in such a way that even for a system with 20\% of disorder and with a misalignment of $10$ degrees, the ground state vertices constitute more than 85\% of the lattice's vertices as shown in Fig.~\ref{theta10} (right). The formation of the second plateau is also mediated by the creation and propagation of type $4$ vertices and may be summarized as follows: $V1 \to V4 \to V5_{xy}$, where $V5_{xy}$ is a type $5$ vertex with magnetization along the diagonal. In the transition between plateaus, $V4$ vertices propagate along diagonal and only spins belonging to group $3$ flip. The recovery of the state magnetized along the $x$-direction can be summarized as: $V5_{xy} \to V6 \to V5_x$, where $V5_{x}$ is a type $5$ vertex with magnetization along $x$-direction. These processes are also shown in the supplementary material\cite{supplementary}.

A very important aspect of our results is that if the external magnetic field is turned off at any point of the hysteresis curve, no spin flip is observed. This means that the process may be stopped at any field value and continued from where it was, allowing one to prepare samples in the ground state or in any intermediary state to make measurements. Size effects were also found to be negligible in the equilibrium properties. We analyzed lattices with linear size ranging from $6$ to $30$ lattice spacings.

Different kinds of disorder were also studied. By considering disorder in the positions, orientations and magnitude of spins alone we found that the results are qualitatively the same obtained by considering the disorder in the switching field values. Moreover, for the same strength of disorder (for instance, 5\% of disorder in positions, orientations, magnitude or switching field), the case with disorder in positions alone have shown to exhibit larger deviations from the perfect system. Then, for a better control over system disorder, the control over the islands positions (or center of mass) seems to be the most important.

The use of an external rotating and decreasing magnetic field, as the one used in demagnetization protocols for the square spin ice~\cite{Wang07}, was also studied. However, we were not able to drive the system to the ground state. Our best result was obtained for a perfect system, where all islands have the same value for the switching field, and 95\% of the vertices were found to be in topology $V1$ at the end of the process. On the other hand, for 10\% of disorder in the switching field, only 45\% of the vertices were found in topology $V1$. Moreover, the final vertex population has no relationship to its energy for high disorder in the system (20\%). It seems in our simulations that the same dynamical bottleneck observed for such a protocol in simulations of the square spin ice~\cite{Budrikis10,Budrikis11,Budrikis12,Budrikis12b,Budrikis12c} is present in the ATSI in such a way that to achieve the ground state a specific path in the phase space must be followed. Nevertheless, we would like to stress that only a small portion of the huge parameter space for such a protocol were investigated until now and a throughout investigation may point towards better demagnetization or even the achievement of the ground-state. 

Our results show that the magnetization changes in the system are mediated by the creation and propagation of monopole-like excitations ($V4$ and $V6$ vertices). However, the monopole picture for the excitations in the ATSI was obtained by considering excitations living in a ground state background. Of course, this picture may not be valid for excitations in general, in special the interaction between excitations that arises in the aligned state (type $4$ vertices in a background of type $5$ vertices, for example) may not have Coulombian contributions. However, applying the same ideas used to study excitations above the ground state in Refs.~\onlinecite{Mol09,Mol10,Mol12} to excitations above the aligned state composed by $V5_x$ vertices, we get almost the same phenomenology with an important difference: the string tension may become negative, an artifact brought about by considering excitations above an excited state. Indeed, such a calculation for a linear string along $x$-direction ($V4$ monopoles and a ``string'' composed by $V1$ vertices)  gives a potential expression $V=q/r+br+c$ with $q = -3.5$, $b = -53$ and $c = 13$. On the other hand, for $V6$ monopoles moving along the diagonal and linked by a $V5_{xy}$ string, we get $q = -3.5$, $b = 1.2 $ and $c = 13$.  Amazing in this case is the fact that the ``monopoles'' charge calculated in this way agrees with the one obtained for excitations above the ground state~\cite{Mol12}. Then, we may speculate that those observations indicate that in a situation where all vertices that satisfy the ``ice rule'' have the same energy, the string may loose its effective tension, as occurs in the crystalline spin ice. Whether this is achieved by thermodynamic effects or others does not matter. Indeed, it may occur that in a system where the population of type $1$, $2$ and $5$ vertices is the same, the effective or mean string tension may go to zero, but a detailed study of this situation is beyond the scope of this paper.

It should be noted that by proper design the mean switching field and inter island interactions can be controlled, changing energy scales and the demagnetization process dynamics. Then, in a lattice of non-interacting islands, i.e., in the limit of large lattice spacing, one may also expect to obtain a $m_x=0$ plateau composed by $V1$ vertices only since islands aligned to the external field are expected to flip first. However, interactions between islands introduce correlations in the reversal process in such a way that it is possible to go from a completely uncorrelated reversal process, similar to the 'popping-noise' behavior observed in Ref.~\onlinecite{Schumann10}, to an avalanche-like behavior as seen in Ref.~\onlinecite{Mengotti10}. Our choice of constants seems to be in an intermediary parameter region, displaying both 'popping-noise' and avalanches behaviors, with clear correlations, mainly near the systems' edges. Then we may expect that for a smaller lattice constant or switching field, inner fields would play a prominent role, increasing correlations and favoring avalanche-like behavior, in such a way that the monopolar character of defects may be easier detected.

In summary, our results point forward the feasibility to drive the artificial spin ice in a triangular lattice to its ground state, even for a relatively strong degree of disorder in the system. To this end, a simple hysteresis-like process may be used to obtain the desired state. In addition, by controlling the angle between the lattice and the external magnetic field, different processes can be obtained allowing a better control over the vertices types present in the re-magnetization transition. We have also observed that transitions are mediated by the creation and propagation of vertices that do not satisfies the $3$-in/$3$-out ice rule and that they can be viewed as magnetic monopole excitations, interacting via Coulombian interactions added by a linear string potential. These results constitute an important step towards a more detailed experimental study of monopole-like excitations in artificial spin ice systems, in special with regard to the experimental determination of the interaction potential between defects. Once this is achieved, an avenue for externally control the magnetic excitations is open, rendering us with the possibility of manipulate the dynamics of magnetic charges, say, a sort of magnetricity in nanostructured frameworks.

\begin{acknowledgments}
We would like to thank J.P. Morgan for helpful comments. The authors thank CNPq, FAPEMIG and CAPES (Brazilian agencies) and PRPq/UFMG for financial support.
\end{acknowledgments}

\bibliography{refs}

\end{document}